\documentclass{iau} 
\usepackage{graphicx}

\title[IAU S343 AGB stars] 
{SWAG: Distribution and Kinematics of an Obscured AGB Population toward the Galactic Center}

\author[J\"urgen Ott et al.]   
{J\"urgen Ott$^{1,2}$, David S. Meier$^{2,1}$, Adam Ginsburg$^{1}$, Farhad Yusef-Zadeh$^{3}$, 
 Nico Krieger$^{4}$, \and Cornelia J\"aschke$^{4}$}

\affiliation{$^1$National Radio Astronomy Observatory, \\ 1003 Lopezville
  Road, Socorro, NM 87801, USA\\ email: {\tt jott@nrao.edu; aginsbur@nrao.edu}
  \\[\affilskip]
$^2$  New Mexico Institute of Mining and Technology, \\ 801 Leroy
Place, Socorro, NM 87801, USA\\ email: {\tt david.meier@nmt.edu} \\[\affilskip]
$^3$  Department of Physics and Astronomy and CIERA, \\Northwestern
University,\\ Evanston, IL 60208, USA\\ email: {\tt
  zadeh@northwestern.edu} \\[\affilskip]
$^4$ Max-Planck-Institut f\"ur Astronomie,\\ K\"onigstuhl 17, 69120
Heidelberg, Germany\\ email: {\tt krieger@mpia.de; jaeschke@mpia.de}}

\pubyear{2018}
\volume{343}  
\jname{Why Galaxies Care About AGB Stars: A Continuing Challenge through Cosmic Time}
\editors{F. Kerschbaum, M. Groenewegen, and H. Olofsson, eds.}
\begin{document}

\maketitle

\begin{abstract}
  Outflows from AGB stars enrich the Galactic environment with metals
  and inject mechanical energy into the ISM. Radio spectroscopy can
  recover both properties through observations of molecular lines. We
  present results from SWAG: ``Survey of Water and Ammonia in the
  Galactic Center''. The survey covers the entire Central Molecular
  Zone (CMZ), the inner $3.35^{\circ}\times0.9^{\circ}$
  ($\sim480\times130$\,pc) of the Milky Way that contains
  $\sim 5\times10^{7}$\,M$_{\odot}$ of molecular gas. Although our survey
  primarily targets the CMZ, we observe across the entire sightline
  through the Milky Way. AGB stars are revealed by their signature
  of double peaked 22\,GHz water maser lines. They are distinguished by
  their spectral signatures and their luminosities, which reach up to
  $10^{-7}$\,L$_{\odot}$. Higher luminosities are usually associated with
  Young Stellar Objects located in CMZ star forming regions. We detect
  a population of $\sim 600$ new water masers that can likely be
  associated with AGB outflows. 

\end{abstract}

\firstsection 

\section{SWAG}

SWAG (``Survey of Water and Ammonia in the Galactic center'') is a
Large Project ($\sim460$\,h on-source) to observe the entire CMZ (inner
500\,pc) of the Milky Way with the Australia Telescope Compact Array
(\cite[Krieger 2017]{kri17}; \cite[J\"aschke 2018]{jae18}; Ott et al. 2018 in
prep.). We target 42 molecular lines and wideband continuum in the
21.2-25.4\,GHz range across $\sim 6500$ individual pointings. The spatial
resolution is $\sim20$'' ($\sim 0.8$\,pc) with up to 0.4\,km\,s$^{-1}$
spectral resolution. The line list contains typical shock,
photon-dominated region, density, and temperature molecular tracers,
as well as radio recombination lines. The list includes the
$6_{16}-5_{23}$ 22\,GHz water (H$_2$O) maser and multiple ammonia lines.

\section{Water Masers toward the CMZ}

H$_2$O masers are typically associated with shocked regions, where
density and temperatures are high enough to pump and invert the water
level populations by collision, and path lengths long enough for
amplification. 22\,GHz is one of the brightest and best accessible
water masers. Maser conditions are typically met in outflows,
frequently in Young Stellar Object (YSOs) jets or envelopes of
Asymptotic Giant Branch (AGB) stars.

\begin{figure}[b]
\begin{center}
 \includegraphics[width=\textwidth]{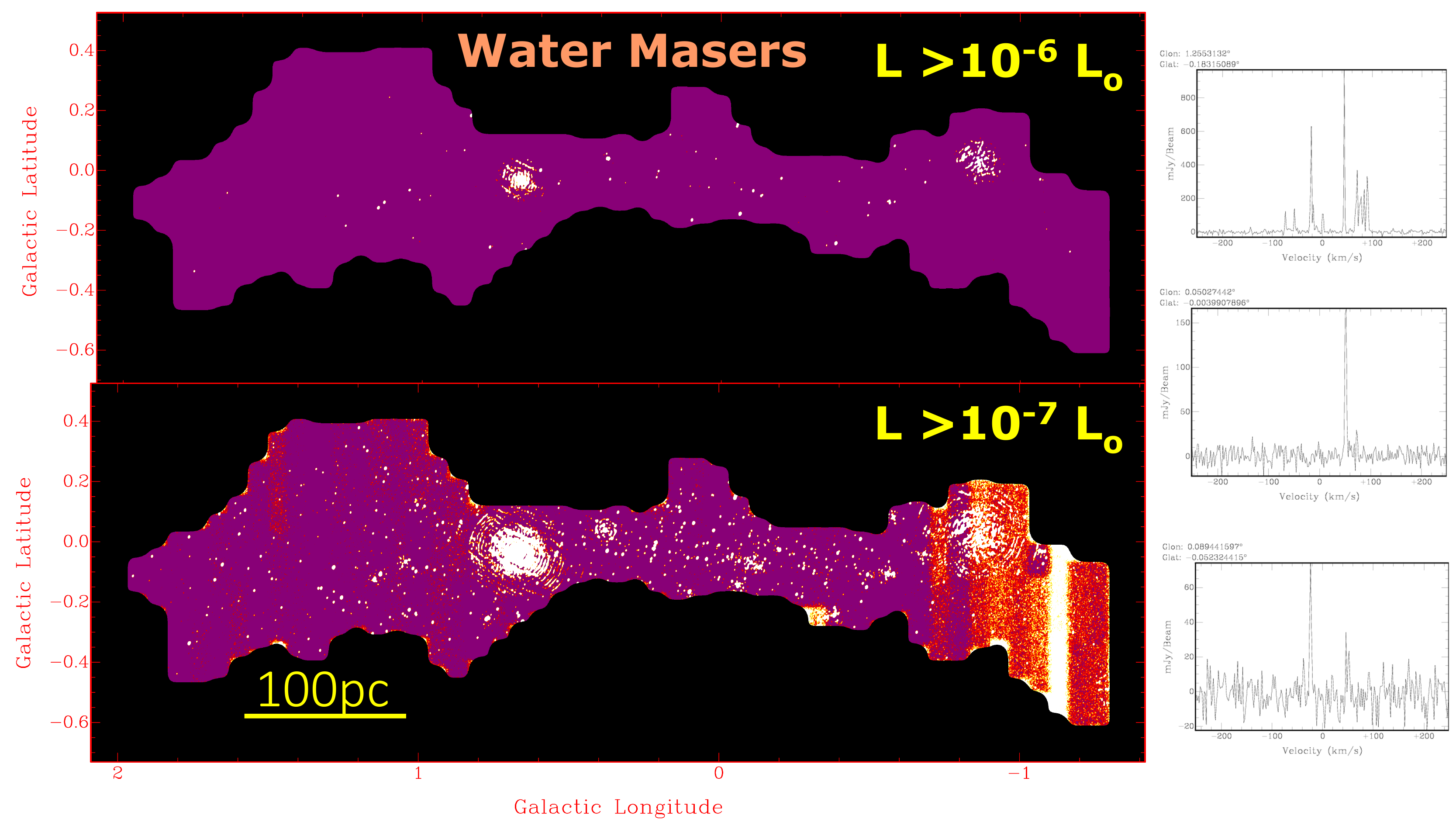} 
 \caption{22\,GHz water masers toward the CMZ. {\bf Top panel:} Masers
   with luminosities $>10^{-6}$\,L$_{\odot}$; {\bf Bottom panel:}
   Masers with $L>10^{-7}$\,L$_{\odot}$. The distribution considerably
   broadens on the faint end. {\bf Right:} Sample spectra for different maser
   luminosities. Faint masers tend to show a second, even weaker
   velocity component, separated by typically $\sim 80$\,km\,s$^{-1}$.}
   \label{fig1}
\end{center}
\end{figure}

The CMZ contains star forming regions and indeed we detect bright
water masers that we associate with YSOs (\cite[Rickert
2017]{ric17}). In the upper panel of Fig.\,\ref{fig1}, we show masers
with luminosities $>10^{-6}$\,L$_{\odot}$ (assuming Galactic Center
distance of 8.5\,kpc for all sources). These masers appear to be
frequently associated with dense molecular gas.

When the luminosity threshold is lowered by an order of magnitude to
$\sim 10^{-7}$\,L$_{\odot}$, the distribution changes and widens in
Galactic Latitude and Longitude. Overall, the masers follow more the
Galactic potential than the distribution of dense gas in the CMZ. In
addition, the weaker masers frequently show a second velocity
component, with an average separation of $\sim 80$\,km\,s$^{-1}$.
High luminosity masers, in contrast, typically show many more
individual components at a greater velocity variation. Example spectra
are shown in Fig.\,\ref{fig1}. The spatial distribution and spectral
signatures of the newly detected, faint population of $\sim 600$
masers (density of $\sim 200$\,deg$^{-2}$) is in agreement with a
population of predominantly AGB stars across the entire Milky Way. The
velocity separation of the maser components suggests that the traced
AGB stars show expansion velocities with typical values around
$\sim 40$\,km\,s$^{-1}$.

\begin{thebibliography}{}

\bibitem[J\"aschke (2018)]{jae18}{J\"aschke, C.} 2018, BS Thesis,
    University of Heidelberg, Germany
\bibitem[Krieger \etal\ (2017)]{kri17}{Krieger, N. \etal}
    \textit{ApJ}, 2017, 850, 77
\bibitem[Rickert (2017)]{ric17}{Rickert, M.} 2017, PhD Thesis,
    University of Illinois, USA




 

\end{thebibliography}
\end{document}